\documentclass[dvips]{article}
\usepackage{graphicx}
\usepackage{subfigure}
\newcommand{\bmpt}[1]{\begin{minipage}[t]{#1\textwidth}}
\newcommand{\bmp}[1]{\begin{minipage}{#1\textwidth}}
\newcommand{\emp}{\end{minipage}}

\newcommand{\ben}{\begin{enumerate}}
\newcommand{\een}{\end{enumerate}}
\newcommand{\be}{\begin{equation}}
\newcommand{\ee}{\end{equation}}
\newcommand{\bea}{\begin{eqnarray}}
\newcommand{\eea}{\end{eqnarray}}
\newcommand{\beaa}{\begin{eqnarray*}}
\newcommand{\eeaa}{\end{eqnarray*}}
\newcommand {\bd}{\begin{description}}
\newcommand {\ed}{\end{description}}
\newcommand{\bi}{\begin{itemize}}
\newcommand{\ei}{\end{itemize}}
\newcommand{\bc}{\begin{center}}
\newcommand{\ec}{\end{center}}
\newcommand {\bq}{\begin{quotation}}
\newcommand {\eq}{\end{quotation}} 
\newcommand {\bdo}{\begin{document}} 
\newcommand {\edo}{\end{document}} 
\newcommand{\ev}[1]{\left<#1\right>}
\newcommand{\eql}{\: = \:}
\newcommand{\dfl}{\: \equiv \:}
\newcommand{\onder}[2]{#1_{\mbox{\scriptsize #2}}}
\newcommand {\boven}[2]{#1^{\mbox{\scriptsize #2}}}
\newcommand{\argmin}[1]{\mathop{\rm argmin}\limits_{#1}}
\newcommand{\klein}[2]{#1_{\mbox{\tiny #2}}}
\newcommand{\doo}{\mbox{do}}
\newcommand{\fracd}[2]{{{\displaystyle #1} \over {\displaystyle #2}}}

\newcommand{\intarg}[2]{\!#2\,d#1}
\newcommand{\del}[3]{\left#1 #3 \right#2}
\newcommand{\av}[1]{\del{<}{>}{#1}}
\newcommand{\bra}[1]{\del{<}{|}{#1}}
\newcommand{\ket}[1]{\del{|}{>}{#1}}
\newcommand{\avmini}[1]{\del{<}{>_{\setminus i}}{#1}}
\newcommand{\avminij}[1]{\del{<}{>_{\setminus i,j}}{#1}}
\newcommand{\sminij}{s_{\setminus i,j}}
\newcommand{\ui}{\avmini{u_i^2}}
\newcommand{\hi}{\avmini{h_i}}
\newcommand{\br}[1]{\del{\{}{\}}{#1}}
\newcommand{\ar}[1]{\del{(}{)}{#1}}
\newcommand{\bl}[1]{\del{[}{]}{#1}}

\newcommand {\Hxx}{H^{(xx)}}
\newcommand {\Hxy}{H^{(xy)}}
\newcommand {\Hyx}{H^{(yx)}}
\newcommand {\Hyy}{H^{(yy)}}
\newcommand {\Cxx}{C^{(xx)}}
\newcommand {\Cxy}{C^{(xy)}}
\newcommand {\Cyx}{C^{(yx)}}
\newcommand {\Cyy}{C^{(yy)}}
\newcommand {\cx}{c^{(x)}}
\newcommand {\hy}{h^{(y)}}
\newcommand{\bt}{\bar{t}}
\newcommand {\bx}{\bar{x}}
\newcommand {\bX}{\bar{X}}
\newcommand {\by}{\bar{y}}
\newcommand {\bw}{\bar{w}}
\newcommand {\bh}{\bar{h}}
\newcommand {\vs}{\vec{s}}
\newcommand {\vbfs}{{\bf s}}
\newcommand {\vv}{\vec{v}}
\newcommand {\vh}{\vec{h}}
\newcommand {\vk}{\vec{k}}
\newcommand {\vz}{\vec{z}}
\newcommand {\vx}{\vec{x}}
\newcommand{\vxv}{\vec{x}_v}
\newcommand{\vxh}{\vec{x}_h}
\newcommand {\vy}{\vec{y}}
\newcommand {\bfy}{{\bf y}}
\newcommand {\bfx}{{\bf x}}
\newcommand {\bft}{{\bf t}}
\newcommand {\bfm}{{\bf m}}
\newcommand {\bfu}{{\bf u}}
\newcommand {\bfw}{{\bf w}}
\newcommand {\bfW}{{\bf W}}
\newcommand {\bfmu}{{\bf \mu}}
\newcommand {\vw}{\vec{w}}
\newcommand {\vu}{\vec{u}}
\newcommand {\vW}{\vec{W}}
\newcommand {\vM}{\vec{M}}
\newcommand {\sign}{\mbox{sign}}
\newcommand {\sech}{\mbox{sech}}
\newcommand {\hmu}{h^{\mu}_j}
\newcommand {\gmu}{g^{\mu}_j}
\newcommand {\gjkmu}{g^{\mu}_{jk}}
\newcommand {\sia}{s_i^\alpha}
\newcommand {\sjb}{s_j^\beta}
\newcommand {\hia}{h_i^\alpha}
\newcommand {\hjb}{h_j^\beta}
\newcommand {\mia}{m_i^\alpha}
\newcommand {\mib}{m_i^\beta}
\newcommand {\mjb}{m_j^\beta}
\newcommand {\wijab}{w_{ij}^{\alpha\beta}}
\newcommand {\chiijab}{\chi_{ij}^{\alpha\beta}}
\newcommand {\Cijab}{C_{ij}^{\alpha\beta}}
\newcommand {\Hijab}{H_{ij}^{\alpha\beta}}
\newcommand {\he}{h_{\mbox{ext}}}
\newcommand {\xia}{x_{i,\alpha}}
\newcommand {\vxk}{\vec{x}_{\mbox{{\small known}}}}
\newcommand {\vxu}{\vec{x}_{\mbox{{\small unknown}}}}
\newcommand {\Pab}{P_{\alpha\beta}}
\newcommand {\dab}{\delta^{\alpha\beta}}
\newcommand {\dij}{\delta_{ij}}
\newcommand {\erf}{\mbox{Erf}}

\title{Path integrals and symmetry breaking for optimal control theory}
\author{H.J. Kappen\\
Radboud University,\\
Nijmegen, the Netherlands}
\begin{document}
%\tableofcontents
\maketitle
\begin{abstract}
This paper considers linear-quadratic control of a non-linear
dynamical system subject to arbitrary cost.  I show that for this class of stochastic
control problems the non-linear Hamilton-Jacobi-Bellman equation can be
transformed into a linear equation. The transformation is similar to
the transformation used to relate the classical Hamilton-Jacobi equation to
the Schr\"odinger equation. 
As a result of the linearity, the usual
backward computation can be replaced by a forward diffusion process,
that can be computed by stochastic integration or by the evaluation of
a path integral. It is shown, how in the deterministic limit the PMP formalism
is recovered. The significance of the path integral approach is that it
forms the basis for a number of efficient computational methods, such
as MC sampling, the Laplace approximation and the variational
approximation. We show the effectiveness of the first two methods in
number of examples. 
Examples are given that show the qualitative difference between
stochastic and deterministic control and the occurrence of symmetry
breaking as a function of the noise.
\end{abstract}

\section{\label{sec:introduction}Introduction}
The problem of optimal control of non-linear systems in the presence of
noise occurs in many areas of science and engineering. Examples are the
control of movement in biological systems, robotics, and financial
investment policies.

In the absence of noise, the optimal control problem can be solved in
two ways: using the Pontryagin Minimum Principle (PMP)
\cite{pontryagin62} which is a pair of ordinary differential equations
that are similar to the Hamilton equations of motion or the
Hamilton-Jacobi-Bellman (HJB) equation which is a partial differential
equation \cite{bellman64}.

In the presence of Wiener noise, the PMP formalism can be generalized
and yields a set of coupled stochastic differential equations, but they
become difficult to solve due to the boundary conditions at initial and
final time (see however \cite{yong_zhou99}).  In contrast, the
inclusion of noise in the HJB framework is mathematically quite
straight-forward. However, the numerical solution of either the deterministic
or stochastic HJB equation is in general difficult
due to the curse of dimensionality. Therefore, one is interested
in efficient methods for solving the HJB equation. The class of
problems considered below allows for such efficient methods.

In section~\ref{sec:linear_HJB}, we
consider the control of an arbitrary non-linear 
dynamical system with arbitrary cost, but with the restriction, that
the control acts linearly on the dynamics and the cost of the control
is quadratic. 
For this class of problems, the non-linear
Hamilton-Jacobi-Bellman equation can be transformed into a linear
equation by a log transformation of the cost-to-go. The transformation stems back to the early days of quantum
mechanics and was first used by Schr\"odinger to relate
the Hamilton-Jacobi formalism to the Schr\"odinger
equation. See section~\ref{discussion} for a further discussion on this point.
The log transform was first used in the context of control theory by
\cite{fleming78} (see also \cite{fleming92}).

Due to the linear description, the usual backward integration in time of
the HJB equation can be replaced by computing expectation values under a
forward diffusion process. This is treated in
section~\ref{sec:forward}. The computation of the expectation value
requires a stochastic integration over trajectories that can be described
by a path integral (section~\ref{sec:path_integral}). This is an integral over all trajectories starting
at $x,t$, weighted by $\exp(-S/\nu)$, where $S$ is the cost of the
path (also know as the Action) and $\nu$ is the size of the noise. 
It has the 
characteristic form of a partition sum and one should therefore
expect that for different values of the noise $\nu$ the control is
qualitatively different, and that symmetry breaking occurs below a 
critical value of $\nu$.

\begin{figure}
\bc 
\includegraphics[width=0.6\textwidth]{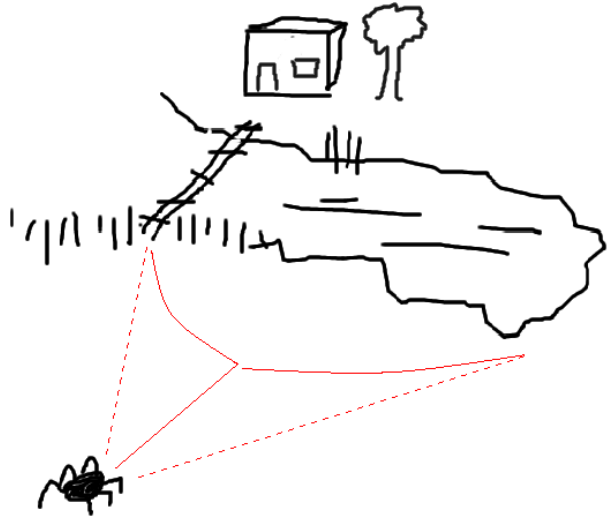}
\ec
\caption{The drunken spider. In the absence of noise (alcohol in this
case), the optimal
trajectory for the spider is to walk over the bridge. When noise is
present, there is a
significant probability to fall off the bridge, incurring a large cost.
Thus, the optimal noisy control is to walk around the lake. }
\label{spider}
\end{figure}

In general, control problems may have several solutions, corresponding to the
different local minima of $S$. 
The case is illustrated in fig.~\ref{spider}. A spider wants to
go home, by either crossing a bridge
or by going around the lake. In the absence of noise, the route over the
bridge is optimal since it is shorter. However, the spider just came out
of the local bar, where it had been drinking heavily with its friends.
He
is not quite sure about the outcome of its actions: any of its movements
may be accompanied by a random sway to the left or right. Since the bridge
is rather narrow, and spiders don't like swimming,
the optimal trajectory is now to walk around the
lake. 
Thus, we see that the optimal control in the presence of noise can be
quantitatively different from the deterministic control. 

In addition to which path to chose, the spider also has the problem {\em
when} to make that decision. Far away from the lake, he is in no position
to chose for the bridge or the detour, as he is still uncertain of where 
his random swaying may bring him. In other words, why would he spend
control effort now to move left or right when there is a 50 \% change that
he may wander there by chance? He decides to delay his choice until he
is closer to the lake. The question is, when should he make his decision
to move left or right?

It is in these multi-modal examples, that the difference between
deterministic and stochastic control becomes most apparent. They are
not only of concern to spiders, but occur quite general in obstacle
avoidance for autonomous systems, differential games, and predator-prey
scenarios. Current efficient approaches to control are essentially
restricted to unimodal situations and therefore cannot address these
issues. The aim of the present paper is to introduce a class of
multimodal control problems that can be efficiently solved using path
integral methods.

The path integral formulation is well-known in statistical
physics and quantum mechanics, and several methods exist to compute
them approximately. The Laplace approximation approximates the
integral by the path of minimal $S$ and is treated in
section~\ref{sec:laplace}. This approximation is exact in
the limit of $\nu\rightarrow 0$, and the deterministic control law is
recovered. The formalism is illustrated for the linear quadratic case
in section~\ref{section:lq}. Further refinements to the Laplace
approximation can be made by
considering the quadratic fluctuations around the deterministic
solution (also know as the semi-classical approximation), but I
believe that this correction has a small effect on the control (it
does strongly affect the value of $J$ but not its gradient). The
semi-classical approximation is not
treated in this paper.

As is shown in section~\ref{sec:multi}, 
in the Laplace approximation 
the optimal stochastic control becomes a mixture of 
deterministic control strategies, weighted by $\exp(-S/\nu)$ and can
be computed efficiently, The path integral
displays a symmetry breaking at a critical value of $\nu$: For large
$\nu$, the optimal control is the average of the deterministic
controls. For small $\nu$, one of the deterministic controls is chosen.
In section \ref{sec:delayed} we give the 
example of the delayed choice problem that displays such symmetry breaking 
as a function of the time to reach the target.

In general, the Laplace approximation may not be sufficiently accurate. 
Possibly the simplest alternative is Monte Carlo (MC) sampling. The naive sampling procedure
proposed by the theory is presented in section~\ref{sec:mc_naive}, but
is shown to be rather inefficient in the double slit example in
section~\ref{section:slit}. It is not difficult
to devise more efficient samplers. In section~\ref{mc_importance}, we propose an
importance sampling scheme, where the sampling distribution is a
(mixture of) diffusion processes with drift given by the Laplace deterministic
trajectories. 
The importance sampling method is compared with the exact results for
the double slit problem in section~\ref{sec:slit_mc}.
In section~\ref{sec:spider}, we compute the optimal
control for the drunken spider for low noise 
using the Laplace approximation and for high noise using 
MC importance sampling.

We begin our story with a brief derivation of the HJB equation for
stochastic optimal control, which is treated in depth in many good
textbooks (see for instance \cite{stengel93, fleming92, yong_zhou99}).

\section{Stochastic optimal control}
\label{sec:stochastic_optimal_control}
Consider the stochastic differential equation
\be
dx=b(x(t),u(t),t)dt+d\xi.
\label{stochastic_dynamics}
\ee
$x,b,d\xi$ and $dx$ are $n$-dimensional vectors and $u$ is an $m$-dimensional vector of controls. $d\xi$ is a Wiener 
processes with $\av{d\xi_k d\xi_l}=\nu_{kl}(x,u,t) dt$.
The initial state of $x$ is fixed: $x(t_i)=x_i$ and the state at final
time $t_f$ is free. 
The problem is to find a control trajectory $u(t),t_i<t<t_f$, such that
\be
C(x_i,t_i,u(\cdot))=\av{\phi(x(t_f))+\int_{t_i}^{t_f}dt
f_0(x(t),u(t),t)}_{x_i}
\label{stochastic_cost}
\ee
is minimal. The subscript $x_i$ on the expectation value is to remind
us that the expectation is over all stochastic trajectories that start
in $x_i$.

The standard construction of the solution for this problem is to set up
a partial differential equation that is to be solved for all times in
the interval $t_i$ to $t_f$ and for all $x$. For this purpose, we define
the {\em optimal cost-to-go function} from any intermediate time $t$
and state $x$:

\bea
J(x,t)&=&\min_{u(t\rightarrow t_f)}C(x,t,u(t\rightarrow
t_f))
\eea
where $u(t\rightarrow t_f)$ denotes the sequence of controls $u(\cdot)$ on the
time interval $[t,t_f]$.
For any intermediate time $t', t<t'<t_f$ we can write a recursive
formula for $J$ in the following way:
\bea
J(x,t)&=&\min_{u(t\rightarrow t_f)}\av{ \phi(x(t_f))+
\int_{t}^{t'}dt f_0(x(t),u(t),t)+\int_{t'}^{t_f}dt
f_0(x(t),u(t),t)}_{x}\nonumber\\
&=&\min_{u(t\rightarrow t')} \av{\int_{t}^{t'}dt f_0(x(t),u(t),t)+
\min_{u(t'\rightarrow t_f)} \av{ \phi(x(t_f))+
\int_{t'}^{t_f}dt f_0(x(t),u(t),t)}_{x(t')}}_{x}\nonumber\\
&=&\min_{u(t\rightarrow t')} \av{\int_{t}^{t'}dt f_0(x(t),u(t),t)+
J(x(t'),t')}_{x}
\label{stochastic_hjbe0}
\eea
The first line is just the definition of $J$. In the second line, we
split the minimization over two intervals. 
These are not independent, because the second minimization is
conditioned on the starting value $x(t')$, which depends on the
outcome of the first minimization. The last line uses again the definition
of $J$.

Setting $t'=t+dt$ we can Taylor expand $J(x(t'),t')$ around $t$.
This expansion takes place within the expectation value and need to be
performed to first order in $dt$ and second order in $dx$, since
$\av{dx^2}={\cal O}(dt)$. This is the standard It\^o calculus argument.
Thus,
\beaa
\av{J(x(t+dt),t+dt)}_{x}&=&\av{J(x,t)+\partial_t
J(x,t)dt+(\partial_x J(x,t))^Tdx+\frac{1}{2}\mathrm{Tr}\left(\partial_x^2
J(x,t)dx^2\right)}\\
&=&
J(x,t)+\partial_t
J(x,t)dt+(\partial_x J(x,t))^T b(x,u,t)dt+\frac{1}{2}\mathrm{Tr}\left(\partial_x^2
J(x,t)\nu(x,u,t)\right) dt
\eeaa
In this expression, $\partial_t$ and $\partial_x$ denotes partial
differentiation with
respect to $t$ and $x$, respectively. Similarly, $\partial^2_x J$ is
the matrix of second derivatives of $J$ and $\mathrm{Tr}(\nu
\partial_x^2 J)=\sum_{ij}\nu_{ij}\frac{\partial^2 J}{\partial x_i
\partial x_j}$.
Substituting this into Eq.~\ref{stochastic_hjbe0}, dividing both sides by $dt$ and
taking the limit of $dt\rightarrow 0$ yields 
\be
-\partial_t J(x,t)=\min_{u}\left( f_0(x,u,t)+b(x,u,t)^T \partial_x J(x,t)+
\frac{1}{2}\mathrm{Tr}\left(\nu(x,u,t) 
\partial_x^2 J(x,t)\right)\right),\quad
\forall t,x
\label{stochastic_hjbe}
\ee
which is the {\em Stochastic Hamilton-Jacobi-Bellman Equation} 
with boundary condition $J(x,t_f)=\phi(x)$. 

Eq.~\ref{stochastic_hjbe} reduces to the deterministic HJB equation in
the limit $\nu\rightarrow 0$. In that case, an alternative approach to
solving the control problem is the Pontryagin Maximum principle (PMP),
which requires the solution of $2n$ ordinary differential equations.
These equations need to be solved with multi-point boundary conditions at
both $t_i$ and $t_f$. Solving $2n$ ordinary differential
equations may be more efficient than solving the $n$-dimensional partial
differential equation, using shooting methods (see for instance
\cite{shampine03}), but may be unstable in some cases.

In the stochastic case, there does not exist a
generic alternative to solving the pde (see however \cite{yong_zhou99} for
stochastic versions of the PMP approach). 
Thus, for stochastic control
one needs to solve the HJB equation, which suffers from the curse of
dimensionality. 

A notable exception is when $b$ is linear in $x$ and $u$ and $f_0$ is
quadratic in $x$ and $u$. This is called the linear-quadratic (LQ)
control problem. In that case, it can be shown that the solution for
$J(x,t)$ is quadratic in $x$ with time-varying coefficients. These
coefficients satisfy coupled ordinary differential (Ricatti) equations that can
be solved efficiently \cite{stengel93}.

\section{A path integral formulation for control}
\subsection{A linear HJB equation} 
\label{sec:linear_HJB}
Consider the special case of Eqs.~\ref{stochastic_dynamics} and~\ref{stochastic_cost} where the dynamic is 
linear in $u$ and the cost is quadratic in $u$:
\bea
dx&=&(b(x,t) + B u)dt + d\xi
\label{easy_dynamics}\\
C(x_i,t_i,u(\cdot))&=&\av{\phi(x(t_f))+\int_{t_i}^{t_f}dt\left(\frac{1}{2}u(t)^T R u(t) +
V(x(t),t)\right)}_{x_i}
\label{easy_cost}
\eea
with $B$ an $n\times m$ matrix and $R$ an $m\times m$ matrix. $B$, $R$ and $\nu$ are independent of $x,u,t$. 
$b$ and $V$ are arbitrary functions of $x$ and $t$ and $\phi$ is an
arbitrary function of $x$. In other words, the system
to be controlled can be arbitrary complex and subject to arbitrary
complex costs. The control instead, is restricted to the simple LQ
form.

The stochastic HJB equation~\ref{stochastic_hjbe} becomes
\beaa
-\partial_t J&=&\min_{u}\left( \frac{1}{2}u^T R u + 
V+(b + B u)^T \partial_x J+\frac{1}{2}
\mathrm{Tr}\left(\nu 
\partial_x^2 J\right)\right)
\eeaa
Minimization with respect to $u$ yields:
\be
u=-R^{-1}B^T\partial_x J(x,t)
\label{u}
\ee
which defines the optimal control $u$ for each $x,t$. The HJB equation
becomes
\beaa
-\partial_t J&=&-\frac{1}{2}(\partial_x J)^T B R^{-1} B^T \partial_x J 
+V +b^T \partial_x J + \frac{1}{2}\mathrm{Tr}\left(\nu 
\partial_x^2 J\right)
\eeaa
This partial differential equation must be solved with boundary
condition $J(x,t_f)=\phi(x)$.  Note, that after performing the
minimization with respect to to $u$, the HJB equation has become non-linear in
$J$. 

We can remove the non-linearity and this will turn out to greatly
help us to solve the HJB equation. 
Define $\psi(x,t)$ through $J(x,t)=-\lambda \log \psi(x,t)$, with
$\lambda$ a constant to be defined. Then
\beaa
&&-\frac{1}{2}(\partial_x
J)^T B R^{-1} B^T \partial_x J +\frac{1}{2}\mathrm{Tr}\left(\nu
\partial^2_x J\right)\\
&=&-\frac{\lambda^2}{2\psi^2}\sum_{ij}(\partial_x
\psi)_i (B R^{-1} B^T)_{ij} (\partial_x \psi)_j
+\frac{\lambda}{2\psi^2}\sum_{ij}\nu_{ij}(\partial_x
\psi)_i(\partial_x
\psi)_j-\frac{\lambda}{2\psi}\sum_{ij}\nu_{ij}\frac{\partial^2
\psi}{\partial x_i \partial x_j}
\eeaa
The terms quadratic in $\psi$ vanish if and only if
there exists a scalar $\lambda$ such that
\be
\nu=\lambda B R^{-1} B^T
\label{noise_condition}
\ee
In other words, the matrices $\nu$ and $B R^{-1} B^T$ must be proportional
to each other with proportionality constant $\lambda$. In the one dimensional case, such a
$\lambda$ always exists, and Eq.~\ref{noise_condition} is not a restriction. In
the higher dimensional case, Eq.~\ref{noise_condition} restricts the possible choices for
the matrices $R$ and $\nu$. To get an intuition for this restriction,
consider the case that $u$ and $x$ have the same dimension, $B$
is the identity matrix and both $R$
and $\nu$ are diagonal matrices. Then Eq.~\ref{noise_condition} states $R\propto \nu^{-1}$.
In a direction with low
noise, control is expensive ($R_{ii}$ large)
and only small control
steps are permitted. In the limiting case of no noise, we deduce that
$u$ should be set to zero: no control is allowed in noiseless
directions. In noisy directions the reverse is true: control
is cheap and large
control values are permitted.
Loosely speaking,
Eq.~\ref{noise_condition} states that noise and control should operate
in the same dimensions.
\footnote{
As a natural example, consider a one-dimensional
second order system subject to additive control
$\ddot{\theta}=f(\theta,t)+u$.
The first order formulation is obtained by setting $x_1=\theta$ and
$x_2=\dot{\theta}$. Then
\beaa
dx_i=(b_i(x,t)+B_iu )dt,\quad i=1,2
\eeaa
with $b_1(x,t)=x_2$, $b_2(x,t)=f(x_1,t)$ and $B=(0,1)^T$. Since $u$ is
one-dimensional, $R$ is a scalar and 
\[
B R^{-1} B^T=\frac{1}{R}\left(\begin{tabular}{cc}0 & 0\\0 &
1\end{tabular}\right)
\]
Condition Eq.~\ref{noise_condition} states that the stochastic
dynamics must have the noise restricted to the second component only:
\beaa
dx_i=(b_i(x,t)+B_iu )dt+d\xi\delta_{i,2},\quad i=1,2
\eeaa
with $\av{d\xi^2}=\nu dt$ and $\lambda=\nu R$. 
}

When Eq.~\ref{noise_condition} holds, the quadratic terms in the HJB
equation cancel and the HJB becomes
\bea
\partial_t \psi &=&\left(
\frac{V}{\lambda}
-b^T\partial_x -\frac{1}{2}\mathrm{Tr}(\nu \partial^2_x)\right)
\psi\nonumber\\
&=& -H \psi
\label{hjb_linear}
\eea
with $H$ a linear operator acting on the function $\psi$.
Eq.~\ref{hjb_linear} must be solved backwards in time with
$\psi(x,t_f)=\exp(-\phi(x)/\lambda)$.  However, the linearity allows us
to reverse the direction of computation, replacing it by a diffusion
process, as we will explain in the next section.

To simplify the exposure in the subsequent sections, we assume the control dimension
$m=n$ and $B$ the unit matrix. 

\subsection{Forward diffusion}
\label{sec:forward}
For real functions $\rho$ and $\psi$, define the inner product
$\av{\rho|\psi}=\int dx \rho(x,t) \psi(x,t)$.
Then we can define $H^\dagger$, the Hermitian conjugate of the operator
$H$, with
respect to this inner product as follows.
\beaa
\av{H^\dagger \rho |\psi}&=&\av{\rho|H \psi}=\int dx \rho(x,t) \left(
-\frac{V(x,t)}{\lambda}
+b(x,t) \partial_x
+\frac{1}{2}\sum_{ij}\nu_{ij}\frac{\partial^2
}{\partial x_i \partial x_j}\right)\psi(x,t)\\
&=& \int dx \left(
-\frac{V(x,t)}{\lambda}\rho(x,t)
-\partial_x (b(x,t)\rho(x,t))
+\frac{1}{2}\sum_{ij}\nu_{ij}\frac{\partial^2
}{\partial x_i \partial x_j}\rho(x,t)\right)\psi(x,t) 
\eeaa
where we have performed integration by parts and assume that $\rho$
vanishes at $|x|\rightarrow \infty$. 
Thus, 
\beaa
H^\dagger\rho &=& -\frac{V(x,t)}{\lambda}\rho(x,t)
-\partial_x (b(x,t)\rho(x,t))
+\frac{1}{2}\sum_{ij}\nu_{ij}\frac{\partial^2
}{\partial x_i \partial x_j}\rho(x,t).
\eeaa
Let $\rho(y,\tau|x,t)$ be a probability density, initialized at $t,x$, that evolves forward
in time
according to the diffusion process
\be
\partial_t \rho = H^\dagger \rho
\label{fp}
\ee
with drift $b(x,t)dt$ and diffusion $d\xi$, and with
an extra term due to the potential $V$. Whereas the other two terms
conserve probability density, the potential term takes out probability
density at
a rate $V(x,t)dt/\lambda$. Therefore,
the stochastic simulation of Eq.~\ref{fp} is 
a
diffusion that runs in parallel with the annihilation process:
\bea dx&=&b(x,t) dt + d\xi\nonumber\\
x&=&x+dx, \quad \mbox{with probability}~~ 1-V(x,t)dt/\lambda\nonumber\\
x_i&=&\dagger, \quad \mbox{with probability}~~ V(x,t)dt/\lambda
\label{diffusion} \eea 
where $\dagger$ denotes that the particle is
taken out of the simulation.  
Note that when $V=0$ this diffusion process is
identical to the original control dynamics Eq.~\ref{easy_dynamics} in the
absence of control ($u=0$).

Since $\psi$ evolves backwards in time according to $H$ and $\rho$ evolves
forwards in time according to $H^\dagger$ the inner product $\int dy
\rho(y,\tau|x,t)\psi(y,\tau)$ is time invariant (independent of $\tau$).
Since $\rho(y,t|x,t)=\delta(y-x)$,
it immediately follows that
\bea
\psi(x,t)&=&
\int dy \rho(y,t_f|x,t) \psi(y,t_f)
\label{forward}
\eea
We arrive at the important conclusion that $\psi(x,t)$ can be
computed either by backward integration using Eq.~\ref{hjb_linear} or by forward
integration of a diffusion process given by Eq.~\ref{fp}.
The optimal cost-to-go is finally given by
\bea
J(x,t) &=&
-\lambda \log \int dy \rho(y,t_f|x,t) \exp(-\phi(y)/\lambda)
\label{forward1}
\eea
with $\rho(y,t_f|x,t)$ given by the stochastic process
Eq.~\ref{diffusion}. The optimal control is given by Eq.~\ref{u}.
See section~\ref{section:lq} for a simple Gaussian example that
illustrate these ideas.

\subsection{The path integral formulation}
\label{sec:path_integral}
In this section, we will write the diffusion kernel $\rho(y,t_f|x,t)$
in Eq.~\ref{forward1} as a path integral. 
For an infinitesimal time step $\epsilon$, we can write
the probability to go from $x$ to $y$ 
as an integral over all noise realizations.
The probability of the Wiener
is Gaussian with mean zero and variance $\nu
\epsilon$. 
\iffalse
\beaa
\rho(y,t+\epsilon|x,t)&=&
\int d\xi_1\ldots d\xi_n
\frac{1}{Z_\xi}\exp\left(-\frac{1}{2\epsilon}\xi^T
\nu^{-1} \xi\right) \prod_i \delta(y_i-x_i-b_i(x,t)\epsilon-\xi_i)\\
&=&\frac{1}{Z_\xi}\exp\left(-\frac{1}{2 \epsilon}(y-x-b(x,t)\epsilon)^T
\nu^{-1}(y-x-b(x,t)\epsilon)\right)
\eeaa
with
\beaa
Z_\xi&=&(2\pi \epsilon)^{n/2}\sqrt{\det \nu}=\int d\xi
\exp\left(-\frac{1}{2 \epsilon}\xi^T \nu^{-1} \xi)\right)
\eeaa
\fi
The particle annihilation destroys probability with rate
$V(x,t)\epsilon/\lambda$.
Combining
annihilation with diffusion, we obtain
\beaa
\rho(y,t+\epsilon|x,t)&\propto&\exp\left(-\frac{\epsilon}{\lambda}\left[
\frac{1}{2}\left(\frac{y-x}{\epsilon}-b(x,t)\right)^T
R\left(\frac{y-x}{\epsilon}-b(x,t)\right)+V(x,t)\right]\right)
\eeaa
where we have used $\nu^{-1}=R/\lambda$.

We can write the transition probability as a
product of $n$ infinitesimal transition probabilities:
\beaa
\rho(y,t_f|x,t)&\propto&
\int dx_1\ldots dx_{n-1}\\
&&\exp\left(-\frac{\epsilon}{\lambda}\sum_{i=0}^{n-1}\left[\frac{1}{2}\left(\frac{x_{i+1}-x_i}{\epsilon}-b(x_i,t_i)\right)^T R\left(\frac{x_{i+1}-x_i}{\epsilon}-b(x_i,t_i)\right)+V(x_i,t_i)\right]\right)
\eeaa
In the limit of $\epsilon \rightarrow 0$, the sum in the exponent becomes
an integral: $\epsilon \sum_{i=0}^{n-1}\rightarrow \int_t^{t_f}d\tau$ and thus
we can formally write
\bea
\rho(y,t_f|x,t)
&=& \int [dx]_x^y \exp
\left(-\frac{1}{\lambda}S_\mathrm{path}(x(t\rightarrow
t_f))\right)\label{path_integral1}\\
S_\mathrm{path}(x(t\rightarrow t_f))&=&\int_t^{t_f}
d\tau\left(\frac{1}{2}\left(\frac{dx(\tau)}{d\tau}-b(x(\tau),\tau)\right)^T
R \left(\frac{dx(\tau)}{d\tau}-b(x(\tau),\tau)\right)+V(x(\tau),\tau)\right)
\label{path_action}
\eea
with $x(t\rightarrow t_f)$ a path with $x(\tau=t)=x, x(\tau=t_f)=y$,
$\int [dx]_x^y$ an integral over paths that start at $x$ and end at $y$.
\footnote{
The paths are continuous but non-differential and there are different 
forward are backward derivatives \cite{nelson67,guerra95}. 
Therefore, the continuous time
description of the path integral and in particular $\dot{x}$ 
are best viewed as a shorthand for its finite $n$
description.}

Substituting Eq.~\ref{path_integral1} in Eq.~\ref{forward1} we can absorb
the integration over $y$ in the path integral and find
\bea
J(x,t)&=&
-\lambda \log \int [dx]_x \exp\left(-\frac{1}{\lambda}S(x(t\rightarrow t_f))\right)
\label{path_integral}
\eea
where the path integral $ \int [dx]_x$ is over all trajectories starting at $x$ 
and 
\bea
S(x(t\rightarrow t_f))
&=&\phi(x(t_f))+S_\mathrm{path}(x(t\rightarrow t_f))
\label{action}
\eea
is the Action associated with a path.

The path integral Eq.~\ref{path_integral} is a log partition sum and
therefore can be interpreted as a free energy. 
The partition sum is not over configurations, but over
trajectories. $S(x(t\rightarrow t_f))$ plays the role of the energy of a
trajectory and $\lambda$ is the temperature. 
This link between stochastic optimal control and a free energy has two
immediate consequences. 1) Phenomena that allow for a free energy description,
typically display phase transitions and spontaneous symmetry breaking.
What is the meaning of these phenomena for optimal control? 
2) Since the path integral appears in other
branches of physics, such as statistical mechanics and quantum
mechanics, we can borrow approximation methods from those fields to
compute the optimal control approximately. 
First we discuss the small noise
limit, where we can use the Laplace approximation to recover the PMP
formalism for deterministic control. Also, the path integral shows us how
we can obtain a number of approximate methods: 1) one can combine 
multiple deterministic trajectories to compute the
optimal stochastic control 2) one can use a variational method, replacing
the intractable sum by a tractable sum over a variational distribution and
3) one can design improvements to the naive MC sampling.

\section{The Laplace approximation}
\label{sec:laplace}
\subsection{The Laplace approximation}
When $\lambda$ is small (i.e. $\nu$ is small), we can expand an arbitrary path $\tilde{x}(\tau)$ around the 
classical path:
\[
\tilde{x}(\tau)=x(\tau)+\delta(\tau),\quad t<\tau<t_f
\]
where $x(\tau)$ is the classical path that we need to
determine, and $\delta(\tau)$ is an independent fluctuation of the
path at time $\tau$. Fluctuations are also allowed at
$\tau=t$ and $\tau=t_f$.
The Action Eq.~\ref{action} can be expanded to first order
in $\delta(\tau)$
as
\bea
S(\tilde{x}(t\rightarrow t_f))
&=&S(x(t\rightarrow t_f))
+\delta_i(t_f) \partial_i \phi(x(t_f))\nonumber\\&+&\int_t^{t_f} 
d\tau\left((\dot{x}(\tau)-b(x,\tau))_i R_{ij}
\left(\frac{d}{d\tau}\delta_j(\tau)-\delta_k(\tau) \partial_k
b_j(x,\tau)\right) +\delta_i(\tau) \partial_i
V(x(\tau),\tau)\right)\nonumber\\
%&=&S(x(t\rightarrow t_f))+\delta_i(t_f) \partial_i \phi(x(t_f))\\
%&+&(\dot{x}(t_f)-b(x(t_f),t_f))_i R_{ij} 
% \delta_j(t_f)-(\dot{x}(t)-b(x(t),t))_i R_{ij}
%\delta_j(t)\\
%&-&\int_t^{t_f} d\tau\delta_k(\tau)
%\left(\frac{d}{d\tau}(\dot{x}(\tau)
%-b(x,\tau))_i R_{ik} 
%+(\dot{x}(\tau) -b(x,\tau))_i R_{ij}
%\partial_k b_j(x,\tau)
%-\partial_k V(x(\tau),\tau)\right)\\
&=&S(x(t\rightarrow t_f))+\delta_i(t_f) \left(\partial_i \phi(x(t_f))
+p_j(t_f)\right)-p_j(t) \delta_j(t)\nonumber\\
&-&\int_t^{t_f} d\tau\delta_k(\tau) \left(\frac{d}{d\tau}p_k(\tau)
+p_j(\tau) \partial_k b_j(x,\tau)
-\partial_k V(x(\tau),\tau)\right)
\label{laplace0}
\eea
where $\partial_k$ means partial differentiation with respect to $x_k$,
repeated indices are summed over and $p$ is defined as
\be
p_k(t)=(\dot{x}(t) -b(x,t))_j R_{jk}
\label{laplace1}
\ee
The term proportional to $\delta_k(\tau)$ under the integral must be zero
and defines an ODE for the
classical trajectory:
\bea
\frac{d}{dt}p_k(t)
+\frac{\partial}{\partial x_k}\left(p_j(t) b_j(x,t)
-V(x,t)\right)=0
\label{laplace2}
\eea

Eq.~\ref{laplace1} can be seen as a definition of $p$, but also as 
a dynamical equation for $x$ that must be solved together with the
dynamical equation for $p$, Eq.~\ref{laplace2}. These equations must be
solved with boundary conditions. The boundary condition for $x$ is given
at initial time and
the term proportional to $\delta_i(t_f)$ defines the boundary condition for
$p(t)$ at $t=t_f$:
\bea
x_i(t)=x,\qquad p_j(t_f)=-\frac{\partial\phi(x(t_f))}{\partial x_j}\label{laplace3}
\label{boundary}
\eea

Define the Hamiltonian,
\be
H(x,p,t)=\frac{1}{2}p^T R^{-1}p+p^T b(x,t)-V(x,t)
\label{hamiltonian}
\ee
Then, Eqs.~\ref{laplace1} and~\ref{laplace2} can be written as
\be
\frac{dx}{dt}=\frac{\partial H(x,p,t)}{\partial p},\qquad
\frac{dp}{dt}=-\frac{\partial H(x,p,t)}{\partial x}
\label{pmp}
\ee 
The Hamiltonian system Eqs.~\ref{pmp} with the mixed boundary conditions
Eqs.~\ref{boundary} are the well-known ordinary differential equations
of the Pontryagin Maximum Principle. 

In the Laplace approximation, the path integral Eq.~\ref{path_integral} is 
replaced by the classical trajectory only. Thus,
\beaa
J(t,x)\approx S(x(t\rightarrow t_f))
\eeaa
since fluctuations at initial time are zero: $\delta_i(t)=0$.
The optimal control is given by 
\be u=-R^{-1}\partial_x J\approx -R^{-1}\frac{\delta S(x(t\rightarrow
t_f))}{\delta x(t)}=R^{-1}p(t)=\dot{x}(t)-b(x(t),t)
\label{laplace4} \ee
where we have used $\frac{\delta S(x(t\rightarrow
t_f))}{\delta x(t)}=-p(t)$ from Eq.~\ref{laplace0}.
The intuition of the Laplace approximation is that one needs to solve the
deterministic equations for the whole interval $[t,t_f]$, starting at the
current place $x$. In particular, the end boundary condition (the location
of the target) will affect the location of the optimal path for all
$[t\rightarrow t_f]$. The control is then 
given by the value of the pseudo-gradient $\dot{x}(t)-b(x(t),t)$ on this trajectory.

Note the minus sign in front of $V$ in Eq.~\ref{hamiltonian}, which has the opposite sign from a
normal classical mechanical system. The term $\frac{1}{2}p^T R^{-1}p$ can
be interpreted as the kinetic energy of the system. 
Thus, the 'energy' $H$ is not the sum,
but the difference of kinetic and potential energy. When $H$ does not explicitly depend on
time ($b(x,t)=b(x)$ and $V(x,t)=V(x)$), 
$H$ is conserved under the deterministic control dynamics:
\[
\frac{dH}{dt}=\frac{\partial H}{\partial
x}\frac{dx}{dt}+\frac{\partial H}{\partial p}\frac{dp}{dt}=0
\]
because of Eqs.~\ref{pmp}. To understand this behavior, consider $b=0$. 
Then along the trajectory:
\[
\frac{1}{2}u^T R u=V(x)+H
\]
with $H$ independent of time.
This relation states that the optimal trajectory is such that much
control is spent in areas of large cost and little control is spent in
areas of low cost.

Note, that the optimal control
is independent of the noise $\nu$ as we expect from the Laplace approximation.
Numerically, we can compute the classical trajectory by 
discretizing $x_\mathrm{cl}(\tau)=x_1,\ldots,x_n$ and minimizing $S(x_\mathrm{cl})=S(x_1,\ldots,x_n)$ using a standard minimization method.

\subsection{The linear quadratic case}
\label{section:lq}
To build a bit of intuition for the diffusion process, the 
path integral and Laplace approximation, we 
consider in this section some simple one-dimensional linear quadratic examples.

First consider the simplest case of free diffusion:
\[
V(x,t)=0, \quad b(x,t)=0, \quad \phi(x)=\frac{1}{2}\alpha x^2 \quad
\]
In this case,
the forward diffusion described by Eq.~\ref{fp} and~\ref{diffusion} can be
solved in closed form and is given by a Gaussian with
variance $\sigma^2=\nu (t_f-t)$:
\be
\rho(y,t_f|x,t)=\frac{1}{\sqrt{2\pi}\sigma}\exp\left(-\frac{(y-x)^2}{2\sigma^2}\right)
\label{rholq}
\ee
Since the end cost is quadratic, the optimal cost-to-go Eq.~\ref{forward1} can be computed
exactly as well. The result is
\bea
J(x,t)&=&\nu R \log\left(\frac{\sigma}{\sigma_1}\right) +
\frac{1}{2}\frac{\sigma_1^2}{\sigma^2}\alpha x^2
\label{Jlq}
\eea
with $1/\sigma_1^2=1/\sigma^2+\alpha/\nu R$.
The optimal control is computed from Eq.~\ref{u}:
\[
u=-R^{-1}\partial_x J=-R^{-1}\frac{\sigma_1^2}{\sigma^2}\alpha
x=-\frac{\alpha x}{R +\alpha(t_f-t)}
\label{lqcontrol}
\]
We see that the control attracts $x$ to the origin with a force that
increases with $t$ getting closer to $t_f$. Note, that the optimal control
is independent of the noise $\nu$. This is a general property of LQ
control.

As an extension, we now add a quadratic potential to the above problem:
$V(x)=\frac{1}{2}\mu x^2$.
We now compute the optimal control in the Laplace approximation.
The Hamiltonian is given by Eq.~\ref{hamiltonian}
\[
H(x,p)=\frac{1}{2}R^{-1}p^2-\frac{1}{2}\mu x^2
\]
and the equations of motion and boundary conditions are given by
Eqs.~\ref{pmp} and~\ref{boundary}:
\beaa
\dot{x}&=&p/R\qquad \dot{p}=\mu x\\
x(t)&=&x\qquad p(t_f)=-\alpha x(t_f)
\eeaa
We can write this as the second order system in terms of $x$ only: 
\[
\ddot{x}=\mu x/R,\qquad x(t)=x\qquad \dot{x}(t_f)=-\alpha x(t_f)/R
\]
The solution for $t<\tau<t_f$ is
\[
x(\tau)= A e^{\sqrt{\mu/R}(\tau-t)}+B e^{-\sqrt{\mu/R}(\tau-t)}
\]
The boundary conditions become 
$A+B=x$ and $A \gamma
(\sqrt{\mu/R}+\alpha/R)=B/\gamma(\sqrt{\mu/R}-\alpha/R)$,
$\gamma=e^{\sqrt{\mu/R}(t_f-t)}$ from which we can solve $A$ and $B$. 
The classical Action Eq.~\ref{action} is computed by substituting the
solution for $x$ :
\beaa
S(x(t\rightarrow
t_f))&=&\frac{1}{2}\alpha x(t_f)^2+\frac{1}{2}\int_t^{t_f}d\tau
(R\dot{x}^2(\tau)+\mu x^2(\tau))
=\frac{1}{2}\sqrt{\mu R}x^2\frac{\gamma^2-\frac{\sqrt{\mu
R}-\alpha}{\sqrt{\mu R}+\alpha}}{\gamma^2+\frac{\sqrt{\mu
R}-\alpha}{\sqrt{\mu R}+\alpha}}
\eeaa
which is equal to the cost-to-go in the Laplace approximation. 
The optimal control is minus the gradient of the cost-to-go. 
Note, that the classical trajectory as well as the minimal action only depends on the initial
condition $x$ and the time-to-go $t_f-t$.
For pure diffusion ($\mu\rightarrow 0$) the classical Action reduces to
\[
S(x(t\rightarrow t_f))=\frac{1}{2}\frac{\alpha R x^2}{R + \alpha
(t_f-t)}
\]
which is identical to the exact expression Eq.~\ref{Jlq} except for the
volume factor (which does not affect the control, since it does not depend
on $x$).
\subsection{The multi-modal Laplace approximation}
\label{sec:multi}
The Action $S$ in Eq.~\ref{path_integral} may have more than one
local minimum. This is typical for control problems, where "many roads lead to
Rome". 
Let $x_\alpha(t\rightarrow t_f), \alpha=1,\ldots$ denote the different
optimal deterministic trajectories that we compute by minimizing the Action:
\[
x_\alpha(t\rightarrow t_f)=\mathrm{argmin}_{x(t\rightarrow t_f)} S(x(t\rightarrow t_f)),\quad \alpha=1,\ldots
\]
These trajectories all start at the same value
$x$.
In our drunken spider example, there are two trajectories: one is over the bridge and the
other is around the lake.
Then, in the Laplace approximation the path integral
Eq.~\ref{path_integral} is approximated by these local minima
contributions only:
\bea
J(x,t)&\approx&
-\lambda\log \sum_\alpha \exp(-S(x_\alpha(t\rightarrow
t_f)/\lambda)
\label{laplace_multi}
\eea
The Laplace approximation ignores all fluctuations around the mode. 
Although these fluctuations can be quite big, their $x$ dependence is
typically quite
weak and must come from beyond Gaussian corrections. This can be seen from 
the pure LQ case when the Gaussian fluctuation term in Eq.~\ref{Jlq} 
is independent of $x$. In the LQ case, the Laplace approximation for
the control (not for the cost-to-go) coincides with the exact solution. Therefore, for unimodal problems ($S$ has only one
minimum) one can often safely ignore the contribution of fluctuations
to the control. However, for multi-modal problems  these fluctuation
terms may have a strong $\alpha$
dependence (they have in the spider problem) and therefore play an
important role when weighting the different contributions in
Eq.~\ref{laplace_multi}.

The optimal control becomes a soft-max of deterministic strategies
\beaa
u(x,t)&=&-R^{-1} \sum_\alpha w_\alpha \partial_x S(x_\alpha(t\rightarrow
t_f)\\
w_\alpha&=&\frac{e^{-S(x_\alpha(t\rightarrow
t_f)/\lambda}}{\sum_\beta e^{-S(x_\beta(t\rightarrow
t_f)/\lambda}}
\eeaa
where $\nu$ plays the role of the temperature.

\section{MC sampling}
A natural method for computing the optimal control is by stochastic
sampling. However, as is often the case with MC sampling, a naive sampler
such as the one 
based directly on Eqs.~\ref{diffusion} may be very inefficient. In this
section, we show how this naive sampler works and how it can be improved
using importance sampling.

\subsection{Naive MC sampling}
\label{sec:mc_naive}
The stochastic evaluation of Eq.~\ref{forward} consists of running
$N$ times the diffusion process Eq.~\ref{diffusion} from $t$ to
$t_f$ initialized each time at $x(t)=x$. Denote these $N$ trajectories 
by $x_i(t\rightarrow t_f),i=1,\ldots,N$. Then, $\psi(x,t)$ is estimated by
\be
\hat{\psi}(x,t)=
\sum_{i\in \mathrm{alive}}w_i,\qquad
w_i= \frac{1}{N} \exp(-\phi(x_i(t_f))/\lambda)
\label{mc_psi}
\ee
where 'alive' denotes the subset of trajectories
that do not get killed along the way by 
the $\dagger$ operation. Note that, although the sum is typically over
less than $N$ trajectories, the normalization $1/N$ includes all
trajectories in
order to take the annihilation process properly into account.

The computation of $u$ requires the gradient of $\psi(x,t)$ instead of
$\psi$ itself. First note, that when we vary the initial point of a
path $x(t\rightarrow t_f)$ from Eq.~\ref{laplace0}
and~\ref{laplace1} we obtain 
\[
\frac{\delta S(x(t\rightarrow t_f))}{\delta x(t)} =(\dot{x}(t) -b(x,t)) R
\]
Thus combining Eq.~\ref{u} and Eq.~\ref{path_integral}, we obtain
\[
u=\frac{1}{\psi(x,t)}\int [dx]_x (\dot{x}(t) -b(x,t))
\exp(-S/\lambda)
\]
Note, that we can sample $u$ by the same batch of (naive) trajectories. For
each trajectory, the quantity $\dot{x}(t) -b(x,t)$ is proportional to the
realisation of the noise in the initial time $t$: $\dot{x}(t)
-b(x,t)=d\xi_i(t)/dt$. Therefore,
\bea
\hat{u}dt&=&\frac{1}{\hat{\psi}(x,t)}\sum_{i\in \mathrm{alive}}^N w_i
d\xi_i(t)
\label{mc_u}
\eea
with $w_i$ given by Eq.~\ref{mc_psi}. This expression has a particular
intuitive form. The optimal control at time $t$ is obtained by averaging the
initial noise directions of the trajectories $d\xi_i(t)$, weighted by their
success $w_i$ at the final time $t_f$.

\subsection{Importance sampling}
\label{mc_importance}
The sampling procedure as described by Eqs.~\ref{diffusion}
and~\ref{mc_psi} gives an
unbiased estimate of $\psi(x,t)$ but can be quite inefficient. The problem 
is is well known, and one of the simplest procedures for improving the
sampling is by importance sampling. 
For path integrals this works as follows. We replace the diffusion
process that yields $\rho(y,t_f|x,t)$ with Action $S_\mathrm{path}$
(Eqs.~\ref{path_integral1} and~\ref{path_action}) by another diffusion process, that
will yield $\rho'(y,t_f|x,t)$ with corresponding Action $S'_\mathrm{path}$ .
Then, 
\beaa
\psi(x,t)&=&\int
[dx]_x\exp\left(-S_\mathrm{path}/\lambda\right)\exp\left(-\phi/\lambda\right)\\
&=&\int [dx]_x
\exp\left(-S'_\mathrm{path}/\lambda\right)
\exp\left(-(\phi+S_\mathrm{path}-S'_\mathrm{path})/\lambda\right) 
\eeaa
The idea is to chose the diffusion process $\rho'$ 
such as to make the sampling of the path integral as
efficient as possible. 

A suggestion that comes to mind immediately is to use the Laplace
approximation to compute a deterministic control trajectory
$x^*(t\rightarrow t_f)$. From this, compute 
its derivative $\dot{x}^*(t\rightarrow t_f)$
and define a stochastic process 
to sample $\rho'$ according to
\bea dx&=&\dot{x}^*(t) dt + d\xi\nonumber\\
x&=&x+dx, \quad \mbox{with probability}~~ 1-V(x,t)dt/\lambda\nonumber\\
x_i&=&\dagger, \quad \mbox{with probability}~~ V(x,t)dt/\lambda
\label{importance_diffusion} \eea
The Action $S'_\mathrm{path}$ for the Laplace-guided diffusion is given by
Eq.~\ref{path_action} with $b(x(\tau),\tau)=x^*(\tau),t<\tau<t_f$.

The estimators for $\psi$ and $u$ are given again by 
Eqs.~\ref{mc_psi} and~\ref{mc_u}, with the difference that
\bea
w_i&=&\frac{1}{N}\exp
\left(-\left(\phi(x_i(t_f))+
S_\mathrm{path}(x_i(t\rightarrow t_f))-S'_\mathrm{path}(x_i(t\rightarrow
t_f))\right)/\lambda\right)
\label{importance_weights}
\eea
and $x_i(t\rightarrow t_f)$ is a trajectory from the sampling process
Eq.~\ref{importance_diffusion} instead of Eq.~\ref{diffusion}.
We will illustrate the effectiveness of this approach in
section~\ref{section:slit}.

\section{Numerical examples}
In this section, we introduce some simple one-dimensional examples to
illustrate the methods introduced in this paper. 
The first example is a double slit, and is sufficiently simple that 
we can compute the optimal control by forward diffusion in closed form.
We use this example to compare the Monte Carlo and Laplace
approximations to the exact result. 
Using the double slit example, we show how the optimal cost-to-go
undergoes symmetry breaking as a function of the noise and/or some
other characteristics of the problem (in this case the time-to-go).
When the targets are still far in the
future, the optimal control is to 'steer for the middle' and delay the
choice to a later time. 

The second example is similar to the first, except that the slit is now of
finite thickness, allowing the particle to get lost in one of the holes.
When one hole is narrow and the other wide, this illustrates the drunken
spider problem. We use both the Laplace approximation and the the Monte
Carlo importance sampling to compute the optimal control strategy,
for different noise levels.

\subsection{The double slit}
\label{section:slit}
\begin{figure}
\bc
{\subfigure[The double slit]
{\includegraphics[angle=0,
height=0.3\textwidth]{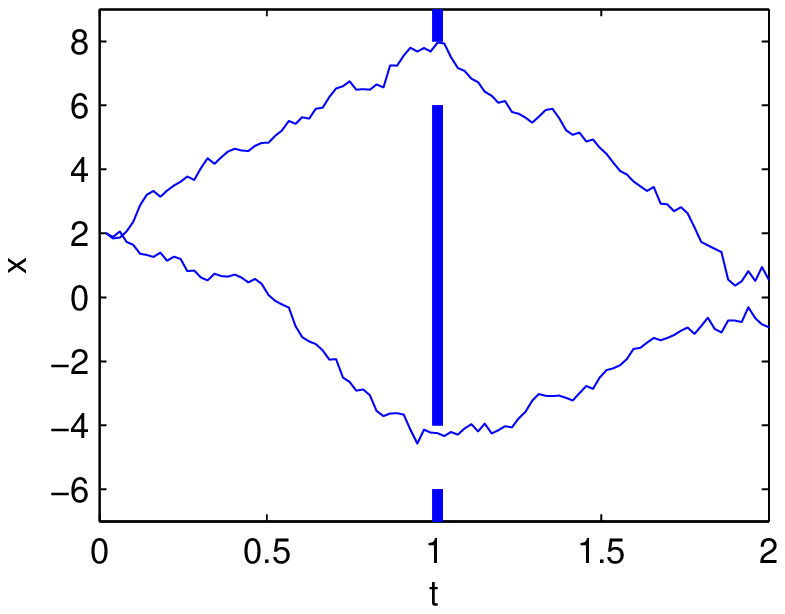}}}\hspace*{5mm}
{\subfigure[Cost-to-go $J(x,t)$]
{\includegraphics[angle=0,
height=0.3\textwidth]{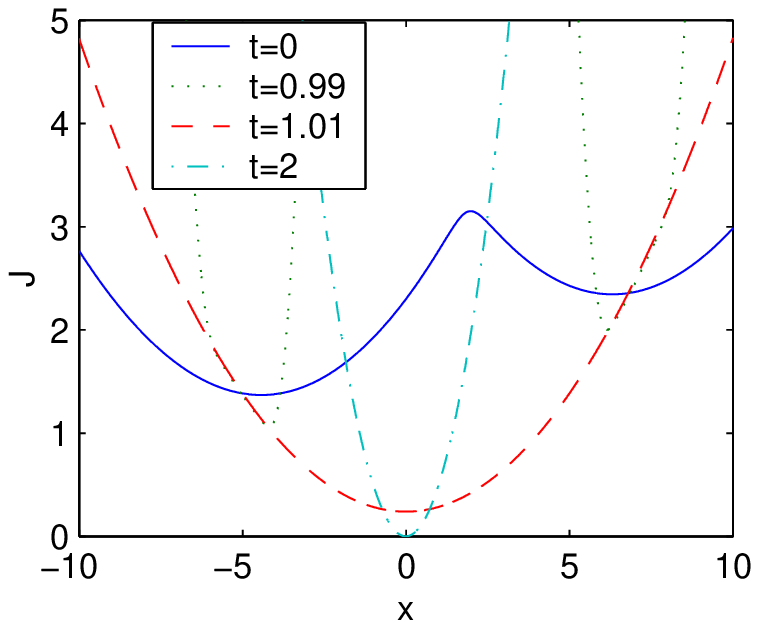}}}
\ec
\caption{(a) The particle moves
horizontally with constant velocity 
from $t=0$ to $t_f=2$ and is deflected up or down by noise and control. 
The end cost $\phi(x)=x^2/2$.
A double slit is placed at $t_1=1$ with openings at $-6<x<-4$ and
$6<x<8$.
Also shown are two example trajectories under optimal control. 
(b) $J(x,t)$ as a function of $x$ for $t=0,0.99,1.01,2$ as computed
from Eq.~\ref{Jlq} and~\ref{slit2}.
$R=0.1, \nu=1, dt=0.02$.
}
\label{slit_file3}
\end{figure}

Consider
a stochastic particle that moves with constant
velocity from $t$ to $t_f$ in the horizontal direction and where
there is deflecting noise in the $x$ direction:
\beaa
dx &=& u dt + d\xi
\eeaa
The cost is given by Eq.~\ref{easy_cost} with
$\phi(x)=\frac{1}{2} x^2$ and
$V(x,t_1)$ implements a slit at an intermediate time $t_1$, $t  <t_1<
t_f$:
\beaa
V(x,t_1)&=&0, \quad a < x < b, \quad c< x < d\\
	&=&\infty, \quad \mathrm{else}
\eeaa
The problem is illustrated in Fig.~\ref{slit_file3}a where the
constant motion is in the $t$ direction and the noise and control is
in the $x$ direction perpendicular to it.

Eq.~\ref{noise_condition} becomes $\lambda=\nu R$ and
the linear HJB becomes:
\[
\partial_t \psi=\left(\frac{V}{\lambda}-\frac{\nu}{2}\partial_x^2\right)\psi
\]
which we must solve with end condition
$\psi(x,t_f)=e^{-\phi(x)/\lambda}$.

Solving this equation by means of the 
forward computation using Eq.~\ref{forward} 
can be done in
closed form. 
First consider the easiest case for times $t>t_1$ where we do not have to
consider the slits. This is the case we have considered before in
section~\ref{section:lq} and the solution is given by Eq.~\ref{Jlq} with
$\alpha=1$.

Secondly, consider $t<t_1$. $\rho(y,t_f|x,t)$ can be written as a
diffusion from $t$ to $t_1$, times a diffusion from $t_1$ to $t_f$
integrating over all $x$ in the slits. 
Substitution in Eq.~\ref{forward} we obtain
\beaa
\psi(x,t)
&=&\int dy \left(\int_a^b +\int_c^d \right)
dx_1\exp(-y^2/2\lambda)\rho(y,t_f|x_1,t_1)\rho(x_1,t_1|x,t)
\eeaa
$\rho(y,t_f|x_1,t_1)$ is Gaussian and given by Eq.~\ref{rholq}. Therefore, we can perform the
integration over $y$ in closed form. We are left with an integral over
$x_1$ that can be expressed in terms of Error functions. The result is
\bea
J(x,t)&=&\nu R \log\left(\frac{\sigma}{\sigma_1}\right) +
\frac{1}{2}\frac{\sigma_1^2}{\sigma^2}x^2-\nu R\log
\frac{1}{2}
\left(F(b,x)-F(a,x)+F(d,x)
-F(c,x)\right)
\label{slit2}
\eea
with
$F(x_0,x)=\erf\left(\sqrt{\frac{A}{2\nu}}(x_0-\frac{B(x)}{A})\right)$,
$A=\frac{1}{t_1-t}+\frac{1}{R+t_f-t_1}$ and $B(x)=\frac{x}{t_1-t}$.
Eqs.~\ref{Jlq} and~\ref{slit2} together provide the solution for the control
problem in terms of $J$ and we can compute the 
optimal control from Eq.~\ref{u}.

A numerical example for the solution for $J(x,t)$ is shown in
fig.~\ref{slit_file3}b. 
The two parts of the solution (compare $t=0.99$ and $t=1.01$) 
are smooth at $t=t_1$ for $x$ in the slits, but
discontinuous at $t=t_1$ outside the slits.
For $t=0$, the cost-to-go $J$ is
higher around the right slit than around the left slit, because the
right slit is further removed from the optimal target $x=0$ and thus
requires more control $u$ and/or its expected target cost $\phi$ is
higher.

\subsubsection{MC sampling}
\label{sec:slit_mc}
\begin{figure}
\bc
{\subfigure[Sample trajectories]
{\includegraphics[angle=0,
height=0.3\textwidth]{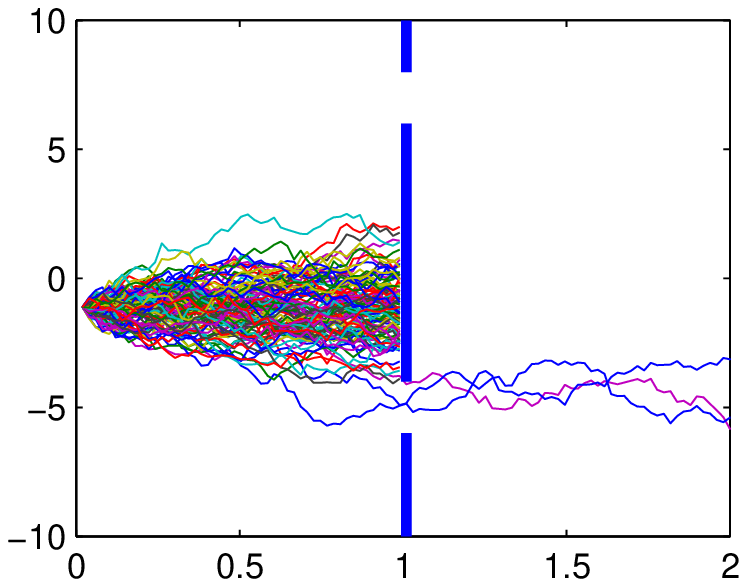}}}\hspace*{5mm}
{\subfigure[MC sampling estimate of $J(x,0)$]
{\includegraphics[angle=0,
height=0.3\textwidth]{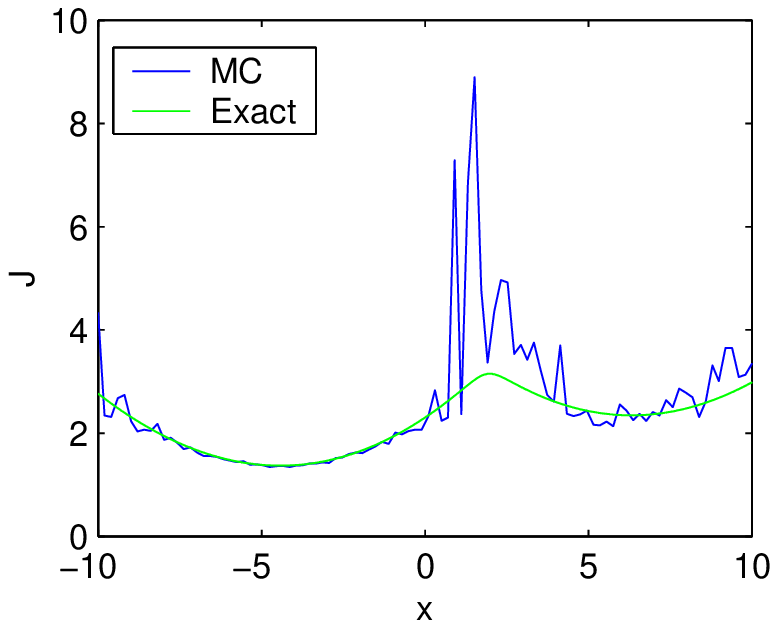}}}
\ec
\caption{Monte Carlo sampling of $J(x,t=0)$
with $\psi$ from Eq.~\ref{diffusion} for the double
slit problem. The parameters are as in fig.~\ref{slit_file3}. (a) Sample of
trajectories that start at $x$ to estimate $J(x,t)$. Only
trajectories that pass through a slit contribute to the estimate.
(b) MC estimate of $J(x,t)=0$ with $N=100000$ trajectories for each
$x$.
}
\label{slit_mcmc2}
\end{figure}

We assess the quality of the naive MC sampling scheme, as
given by Eqs.~\ref{diffusion} and~\ref{mc_psi} in
fig.~\ref{slit_mcmc2}, where we compare $J(x,0)$ as given by
Eq.~\ref{slit2} with the MC estimate Eq.~\ref{mc_psi}.
The left figure shows the trajectories of the sampling procedure for one
particular value of $x$. Note, the inefficiency of the sampler because 
most of the trajectories are killed at the infinite potential at $t=t_1$.
The right figure shows the accuracy of the estimate of $J(x,0)$ for all
$x$ between 
$-10$ and $10$ using $N=100000$ trajectories.
Note, that the number of trajectories that are required to obtain
accurate results, strongly
depends on the value of $x$ and $\lambda$ due to the factor
$\exp(-\phi(x)/\lambda)$ in Eq.~\ref{diffusion}. For high $\lambda$ or low
$\av{\phi}$, few
samples
are required (see the estimates around $x=-4$). For 
small noise or high $\av{\phi}$ the estimate is strongly determined 
by the trajectory with minimal $\phi(x(t_f))$ and many samples may be
required to reach this $x$. In other words, sampling becomes more accurate
for high noise, which is a well-known general feature
of sampling. Also, low values of the cost-to-go are more easy to sample
accurately than high values.
This is in a sense fortunate, since the
objective of the control is to move the particle to lower values of
$J$ so that subsequent estimates become easier.

The sampling is of course particularly difficult in this example because of the
infinite potential that annihilates most of the trajectories. However,
similar effects should be observed in general due to the multi-modality of
the Action. 

\begin{figure}
\bc
\includegraphics[width=0.4\textwidth]{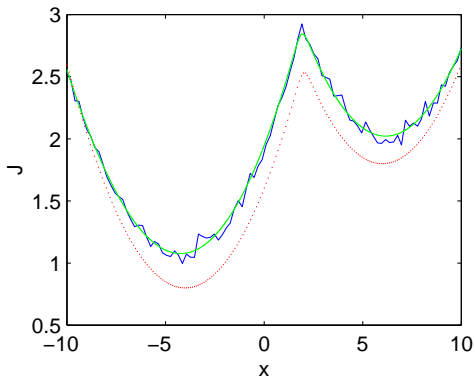}
\ec
\caption{Comparison of
Laplace approximation (dotted line) and Monte Carlo importance sampling
(solid jagged line) of $J(x,t=0)$
with exact result Eq.~\ref{slit2} (solid smooth line) for the double
slit problem. The importance sampler used $N=100$ trajectories for each
$x$. The parameters are as in fig.~\ref{slit_file3}. 
}
\label{slit_mcmc4}
\end{figure}

We can improve the sampling procedure using the importance sampling
procedure outlined in section~\ref{mc_importance}, using the Laplace
approximation. 
The Laplace approximation to $J$ requires the computation of the optimal
deterministic trajectories. In general, one must use some numerical method
to compute the Laplace approximation, for instance minimizing the Action
Eq.~\ref{action} using a time-discretized version of the path. In this
particular example, however, we can just write down the classical
trajectories 'by hand'. For each $x$, there are two
trajectories, each being piecewise linear. 
The Action for each trajectory
is simply 
\[
S_i(x)=\frac{1}{2}R\int_0^2 dt
\dot{x_i}(t)^2=\frac{R}{2}(a_i-x)^2+\frac{R}{2}a_i^2,\quad i=1,2
\]
since
$\phi(x(t_f))=V(x(t_1),t_1)=0$ by construction. $a_i=6$ and $-4$ for the
two trajectories, respectively. The cost-to-go in the Laplace
approximation is given by Eq.~\ref{laplace_multi}:
\[
J_\mathrm{Laplace}(x,0)=-\nu R
\log\left(
\exp\left(-\frac{S_1(x)}{\lambda}\right)+
\exp\left(-\frac{S_2(x)}{\lambda}\right)
\right)
\]
For each $x$, we randomly choose one of the two Laplace
approximations with equal probability. 
We then sample according to Eq.~\ref{importance_diffusion} with
$x^*$ the selected Laplace approximation and estimate $\psi$ using Eq.~\ref{mc_psi} and weights
Eq.~\ref{importance_weights}. 
The Laplace approximation and the results of the importance sampler are given in 
fig.~\ref{slit_mcmc4}.
We see that the Laplace approximation is quite good for this example, in 
particular when one takes into account that a constant shift in $J$ does
not affect the optimal control. The MC importance sampler dramatically
improves over the naive MC results in fig.~\ref{slit_mcmc2}, in particular
since 1000 times less samples are used and is also significantly better
than the Laplace approximation.

\subsubsection{The delayed choice}
\label{sec:delayed}
\begin{figure}
\bc
{\subfigure[Optimal cost-to-go at different $T$.]
{\includegraphics[angle=0,
height=0.3\textwidth]{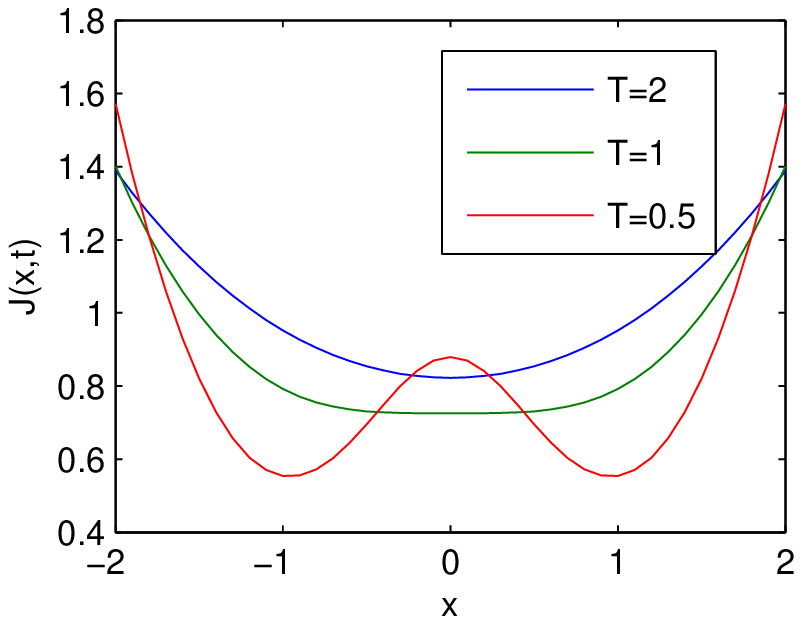}}}
{\subfigure[Sample paths] 
{\includegraphics[angle=0,
height=0.3\textwidth]{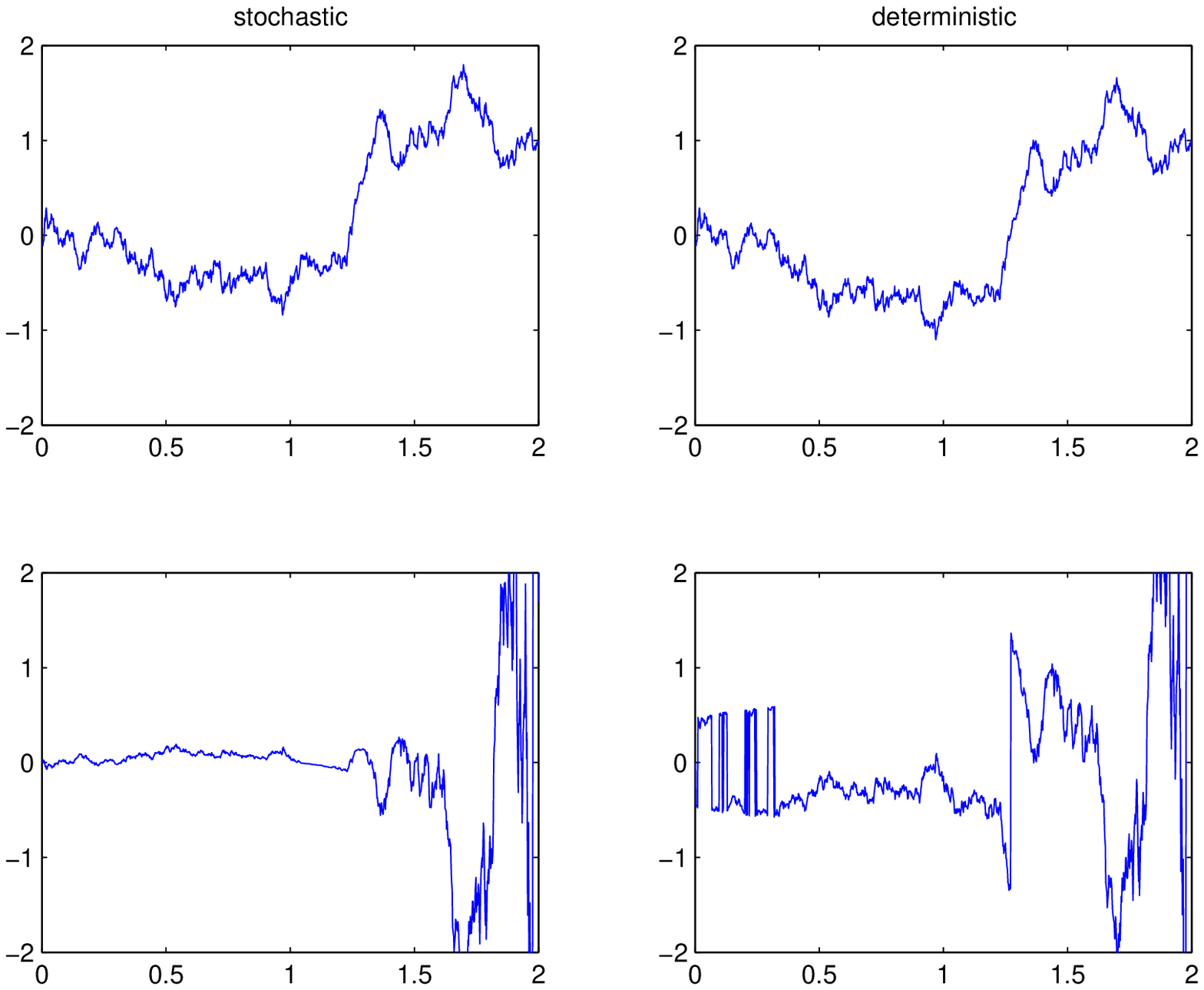}}}
\ec
\caption{
(a) Symmetry breaking in $J$ as a function of $T$ implies a 'delayed
choice'
mechanism for optimal stochastic control.
When the target is far in the future,
the optimal policy is to steer between the targets. Only when $T<1/\nu$
should one aim for one of the targets.
$\nu=R=1$.
(b) Sample trajectories (top row) and controls (bottom row) 
under stochastic control
Eq.~\ref{u_delayed} (left column) and deterministic control
Eq.~\ref{u_delayed} with $\nu=0$ (right column), using identical initial conditions $x(t=0)=0$ and noise
realization.} \label{multi_laplace}

\end{figure}

Finally, we show an example how optimal stochastic control exhibits spontaneous
symmetry breaking. To simplify the mathematics, 
consider the double slit problem, when the size of
the slits becomes infinitesimally small. 
Eq.~\ref{slit2}, with $a=1, b=1+\epsilon,
c=-1-\epsilon, d=-1$ becomes to lowest order in $\epsilon$:
\beaa
J(x,t)&=&
\frac{R}{T}\left(\frac{1}{2}x^2-\nu T \log 2 \cosh\frac{x}{\nu
T}\right)+\mathrm{const.}
\eeaa
where the constant diverges as ${\cal O}(\log \epsilon)$ independent
of $x$ and $T=t_1-t$ the time to reach the slits.
The expression between brackets is a typical free energy with inverse temperature $\beta=1/\nu T$. 
It displays a symmetry breaking at $\nu T= 1$
(fig.~\ref{multi_laplace}a).
For $T>1/\nu$ (far in the past) it is best to steer towards $x=0$ (between the targets)
and delay the choice which slit to aim for until later.  The reason why
this is optimal is that from that position the expected diffusion alone
of size $\nu T$ is likely to reach any of the slits without control
(although it is not clear yet which slit).
Only sufficiently late in time ($T<1/\nu$) should one make
a choice.  
The optimal control is given by the gradient of $J$:
\bea
u=\frac{1}{T}\left(\tanh \frac{x}{\nu T} -x\right)
\label{u_delayed}
\eea

Figure~\ref{multi_laplace}b depicts two trajectories and their
controls under stochastic and deterministic optimal control, using the
same realization of the noise. 
Note, that at early times
the deterministic control drives $x$ away
from zero whereas in the
stochastic control drives $x$ towards zero and smaller in size. 
The stochastic control maintains $x$ around zero and delays the choice
for which slit to aim until $T\approx 1$.

The fact that symmetry breaking occurs in terms of the value of $\nu
T$, is due to the fact that $S\propto 1/T$, which in turn is due to
the fact that $u\propto 1/T$. Clearly, this will not be true in
general. For an arbitrary control problem, $S$ does not need to be
monotonic in $T$, which means that in principle control can be
shifting back and forth several times between the symmetric and the
broken mode as $T$ decreases to zero.

\subsection{The drunken spider}
\label{sec:spider}
\begin{figure}
\bc
{\subfigure[Sample paths for $\nu=0.001$]
{\includegraphics[angle=0, height=0.3\textwidth]{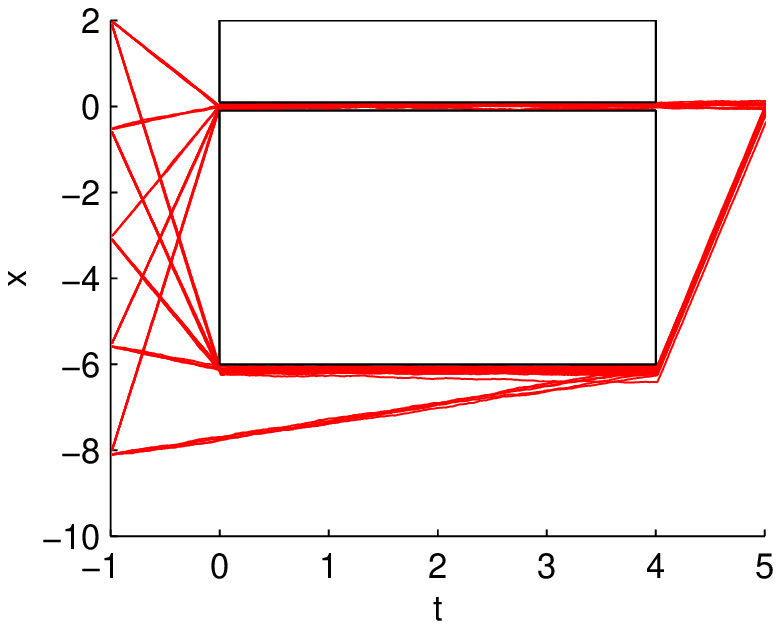}}}
{\subfigure[Sample paths for $\nu=0.1$]
{\includegraphics[angle=0, height=0.3\textwidth]{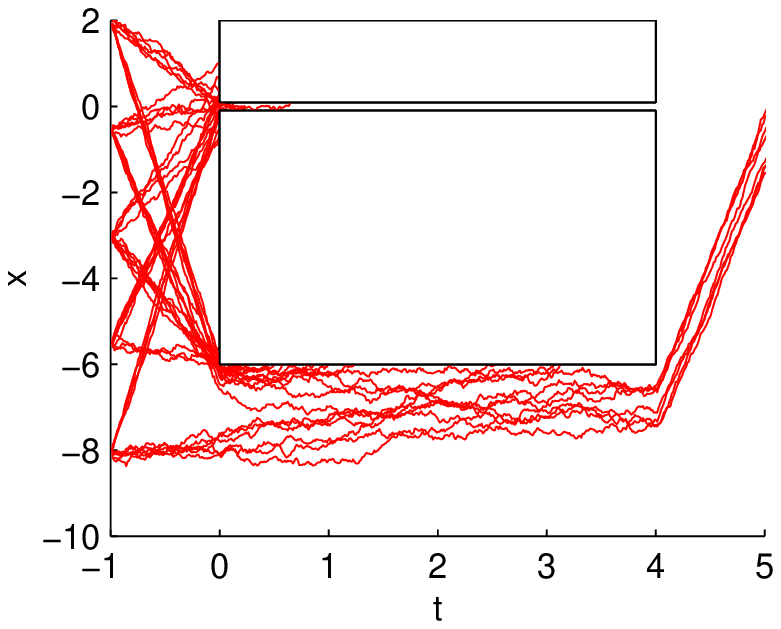}}}
{\subfigure[$J(x,t=-1)$]
{\includegraphics[angle=0, height=0.3\textwidth]{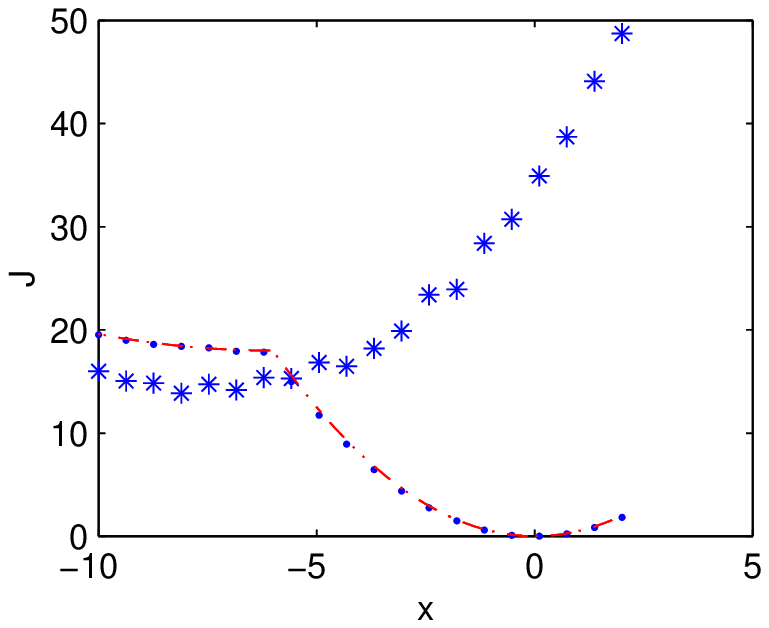}}}\hspace*{5mm}
\ec
\caption{The drunken spider problem. A spider located at $x$ and
$t=-1$ wants to arrive home ($x=0$) at time $t_f$. The lake is
indicated by the white square area, interrupted by a narrow bridge. 
The lake is modelled by the infinite potential given by
Eq.~\ref{lake} with $-a=b=0.1, c=-\infty$ and $d=-6$. $t_1=0, t_2=4,
t_f=5$ and $R=1$. The cost-to-go is computed by
forward importance sampling as outlined in section~\ref{mc_importance}. The guiding Laplace approximations are
the deterministic trajectories over the bridge and around the lake.
Time discretization $dt=0.012$.
(a) Some stochastic trajectories used to compute $J$ for $\nu=0.001$.
(b) Some stochastic trajectories used to compute $J$ for $\nu=0.1$.
(c) The optimal cost-to-go $J(x,t)$ in the Laplace approximation for
$\nu=0.001$ and $\nu=0.1$ solid line (these two curves coincide). The
MC importance sampling estimates are based on $1000$ trajectories per
$x$ for $\nu=0.001$ (dots) and for $\nu=0.1$ (stars). 
}
\label{spider1}
\end{figure}

In order to illustrate the drunken spider problem, 
we change the potential of the double slit problem so that it has a
finite thickness:
$V(x,t)=0$ for all $t<t_1$ and $t>t_2$ and for $t_1<t<t_2$:
\bea
V(x,t)&=&0, \quad a < x < b, \quad c< x < d\nonumber\\
    &=&\infty, \quad \mathrm{else}
\label{lake}
\eea
The problem is illustrated in Fig.~\ref{spider1} and the parameter
values are given in the caption.

The cost-to-go in the Laplace approximation is given by
Eq.~\ref{laplace_multi}, with $S(x(_\alpha(t\rightarrow t_f)),
\alpha=1,2$ the cost of getting home over the bridge or around the lake,
respectively. It is plotted as a function of the current position $x$ 
as the solid line in fig.~\ref{spider1}c, for both $\nu=0.001$ and
$\nu=0.1$ (these two curves coincide for these values of $\nu$, since
$S/\nu$ is so large that the softmax is basically a max). 

In addition, we compute $J$ using importance sampling as outlined
in section~\ref{mc_importance}.  For each $x$, we run $m=1000$
trajectories. For each trajectory, we select randomly one of the
two Laplace trajectories with equal probability, which we denote by
$x^*(t\rightarrow t_f)$.  The stochastic trajectory $x(t\rightarrow t_f)$
is then computed from Eq.~\ref{importance_diffusion}.  It contributes
to the partition sum Eq.~\ref{mc_psi} with a weight that is computed by
Eq.~\ref{importance_weights}, where $S_\mathrm{path}(x(t\rightarrow
t_f))$ and $S'_\mathrm{path}(x(t\rightarrow t_f))$ are
given by Eq.~\ref{path_action} with $b(x(\tau),\tau)=0$ and
$b(x(\tau),\tau)=x^*(\tau)$, respectively.

The results of the MC importance sampling for various $x$ for low noise
($\nu=0.001$) and high noise ($\nu=0.1$) are also shown in 
fig.~\ref{spider1}c.
The dots are the results of the MC
importance sampling at low noise and closely follow the Laplace
results. Note the discontinuous change in slope at $x=-6$, which
implies a discontinuous change in the optimal control value $u$ at that
point: For $x>-6$ the spider steers for the bridge, which requires a
larger control value than for $x<-6$ when the optimal trajectory is 
around the lake. Thus, the optimal path is simply given by the shortest
path and noise is ignored in these considerations.

The MC estimates for $\nu=0.1$ are indicated by the stars in
fig.~\ref{spider1}c. Since noise is large, the Laplace approximation is not
valid, and indeed are very different from the MC estimate.  The
Laplace approximation ignores the effect of deviations from the
deterministic trajectory on the Actions . Thus, it does not take
into account that the spider may wander off the bridge and drowns,
which at this level of noise will happen with almost probability
one and makes $S_\mathrm{bridge}$ much larger than $S_\mathrm{lake}$. 
The MC importance sampling is guided by trajectories around the
lake, that likely survive and by trajectories over the bridge, that
will likely drown and thus will not contribute to Eq.~\ref{mc_psi}.
The estimate for $J$ is thus dominated by trajectories around the lake
and the cost-to-go increases with increasing $x$. Also note, that the
MC estimate puts the minimum of $J$ not at $x=-6$ but safely away from
the lake, so that spider is not likely to fall in the lake on the low
side either, and will have a safe journey home.

\section{Discussion}
\label{discussion}
In this paper, we have addressed the problem of computing stochastic
optimal control.  The direct solution of the HJB equation requires a
discretization of space and time.  This computation naturally becomes
intractable in both memory requirement and cpu time in high dimensions.
We have shown, that for a certain class of problems the control can be
computed by a path integral. The class of problems includes arbitrary
dynamical systems, but with a limited control mechanism. It includes LQ
control as a special case. 
The path integral approach has the advantage that
the $n$-dimensional $x$-space integration of the HJB equation is replaced by an
$n$-dimensional sampling problem. For high-dimensional problems, a
stochastic integration method is expected to be much more efficient
than numerical integration of the HJB equation directly, which scales
exponentially in $n$. 

The obvious approximation methods to use are the Laplace
approximation, the variational approximation and MC sampling. The
Laplace approximation is very efficient. The deterministic
trajectories are found by minimizing the action, which can be done by
standard numerical methods. It typically requires ${\cal
O}(n^2 k^2)$ operations, where $n$ is the dimension of the problem and
$k$ is the number of time discretizations. We have seen that the
multi-modal Laplace approximation gives non-trivial solutions
involving symmetry breaking. 

Computing the path integral by MC sampling is clearly a very generic
approach, that for many practical control applications may well be the
best way to go.  Naive sampling should be replaced by more advanced
sampling schemes. I have only considered one simple improvement using
importance sampling. Other possible improvements could be a Gibbs sampler
or a Metropolis-Hasting sampler.  Clearly, more work in this direction
must be done.

In this paper we have numerically computed the path integrals
using the most simple discretization strategy: short time
averaging \cite{schweitzer_jchemphys81}.  The computation
can be made much more efficient using Fourier discretization
\cite{miller75_jchemphys75,freeman_jchemphys81} or other subspace
approximations (compact splines or wavelets) \cite{bond_jcompphys03}. In
each of these methods the path integral is reduced to a high (but finite)
dimensional Riemann integral, which is approximated using a Monte Carlo
method.  These more advanced discretizations can be combined with any
of the mentioned MC methods.

I have not discussed the variational approximation in this paper. This
approach to approximating the path integral is also known as
variational perturbation theory and gives an expansion of the path
integral in terms of the anharmonic interaction terms and a
variational function that is to be optimized \cite{kleinert04}. The lowest
term in the expansion is similar to what is known as the
variational approximation in machine learning using the Jensen's bound
\cite{feynman_kleinert86}, but one can also consider higher order terms.
The expansion is around a tractable dynamics, such as for instance the
harmonic oscillator, whose variational parameters are optimized such as
to best approximate the path integral.  The application of this method
to optimal control would be the topic of another paper. A complication
of such an analytic treatment is the presence of topological
constraints, such as walls and obstacles.

There exist other fields of research that use path integrals and
where dedicated numerical methods have been developed to solve them.
For instance, in chemical physics path integrals are used to describe
conformational changes in molecules over large time scales. The problem
is similar to an optimal control problem such as navigating a maze: The
begin and end positions are known, and one or more path of minimal cost
needs to be found.  A prominent method in this field is transition path
sampling \cite{bolhuis02}, which can be viewed as a Metropolis-Hasting
sampling scheme in path space, where a new path is sampled by changing
part of the current path and accepting the new path with a probability.
This approach is probably also suitable for optimal control.

There is a superficial relation between the work presented in this paper
and the body of work that seeks to find a particle interpretation
of quantum mechanics. In fact, the $\log$ transformation was
motivated from that work.  Madelung \cite{madelung26} observed that if
$\Psi=\sqrt{\rho}\exp(iJ/\hbar)$ is the wave function that satisfies the
Schr\"odinger equation, $\rho$ and $J$ satisfy two coupled equations. One
equation describes the dynamics of $\rho$ as a Fokker -Planck equation. 
The other equation is a
Hamilton-Jacobi equation for $J$ with an additional term, called the
quantum-mechanical potential which involves $\rho$. Nelson showed that these equations describe
a stochastic dynamics in a force field given by the $\nabla J$, where
the noise is proportional to $\hbar$ \cite{nelson67,guerra81}.

Comparing this to the relation $\Psi=\exp(-J/\lambda)$ used in this
paper, we see that $\lambda$ plays the role of $\hbar$ as in the QM case.
However, the big difference is that there is only one real valued
equation, and not two as in the quantum mechanical case. 
In the control case, $\rho$ is computed {\em as an alternative} to
computing the HJB equation. In the QM case, the dynamics of $\rho$ and $J$ are
computed together.
The QM density evolution is non-linear in $\rho$ because the drift force 
that enters the Fokker-Planck equation depends on 
$\rho$ through $J$ as computed from the HJ equation. 

\section*{Acknowledgement}
I would like to thank Hans Maassen for useful discussions.
I would like to thank Michael Jordan, Peter Bartlett and Stuart Russell
to host my sabbatical at UC Berkeley, which gave me the time to write
this paper. This work is sponsored in part by the Miller Institute for
Basic Research in Science of the University of California at Berkeley
and the ICIS project, grant number BSIK03024.

{\small 
\bibliography{/home/snn/bertk/doc/authors}
\bibliographystyle{unsrt}
}
\end{document}